\documentstyle[prl,twocolumn,aps,floats,epsf,bm]{revtex}
\begin{document}
\twocolumn[
\hsize\textwidth\columnwidth\hsize\csname@twocolumnfalse\endcsname
\draft
\title{Orbital Zeeman effect: Signature of a massive
spin wave mode in ferromagnetism}
\author{Paulo F. Farinas}
\address{Centro de Ci\^ encias Exatas e Tecnol\' ogicas,
Universidade S\~ ao Francisco 13 251-900,
Itatiba, S\~ ao Paulo, Brazil}

\author{Kevin S. Bedell}
\address{Department of Physics, Boston College,
Chestnut Hill, Massachusetts 02167}

\date{\today}
\maketitle

\begin{abstract}
By deriving the quantum hydrodynamic equations
for an isotropic single-band ferromagnet
in an arbitrary magnetic field,
we find that a massive mode recently predicted
splits under the action of the field. 
The splitting is a peculiarity of charged
fermions and is linear in the
field to leading order in $q$ bearing resemblance to the
Zeeman effect in this limit, and providing a clear
signature for the experimental observation of this mode.

\end{abstract}
]
Magnetism in solids has been one of the most studied
subjects in physics for the past decades.
In particular the behavior of propagating spin
waves in metallic materials has now a long
list of contributions in the literature.
Spin waves in ferromagnetic materials
were first predicted by Bloch\cite{blo}
and Slater\cite{sla} and
later observed in iron by Lowde.\cite{low}
These first theories were constructed
based on lattice models of local moments.
Predictions for spin waves using Fermi liquid 
theory\cite{lan} were first made
by Silin for paramagnetic systems.\cite{sil}
Paramagnetic spin-waves were
found to propagate only under an
applied magnetic field.
Abrikosov and Dzyaloshinskii\cite{abr} were the first
to develop a theory of itinerant ferromagnetism based
on Landau's theory of Fermi liquids. Although correct
at the phenomenological level, the microscopic
foundations for this theory were
estabilished only much later
by Dzyaloshinskii and Kondratenko.\cite{dzy}

In this paper we derive the hydrodynamic equations of
the ferromagnetic Fermi liquid theory (FFLT)
for a finite magnetic field and show that
an intrisic degeneracy of a recently predicted
massive mode\cite{bla} exists and is lifted under the
external field, similarly to the Zeeman splitting of
a single spin in a magnetic field. Underneath
this effect we identify the breaking of
chiral symmetry
by the
external magnetic field in the case of charged
fermions, which relates to the orbital motion
developed under the Lorentz force.
Besides being a clear signature of the massive
mode, it is suggested that
quantization of this mode
will lead to a system of
massive magnons with ``up'' and ``down'' states
that split under an external field. We estimate
the values for the fields where the effect
will be observed in typical weak ferromagnets.

For isotropic metals, the equations
describing dynamics of paramagnetic
spin waves are similar to the ones that result from
FFLT
in the small moments limit. This is expected since
in this limit quasi-particles can be defined and one
recovers the kinetic behavior of ordinary Fermi liquid
theory. However, FFLT rests on the assumption of a quite
different, symmetry-broken, ground state and the resulting
spin waves propagate with no external magnetic field
present.
The Goldstone mode associated with spontaneously
broken spin rotation invariance has been the
paradigm of spin-waves
in an isotropic single band ferromagnet
and can be derived from FFLT (as well as from
lattice models).
However, FFLT contains
spin-wave modes that have not yet
been observed, as pointed out
in Ref.\cite{bla}, where the proper hydrodynamics
and parameters for the propagation of the lowest
in energy of these modes have been studied.
Such mode is not a Goldstone mode, hence
its dispersion may in principle be gapped for
$q = 0$. Indeed, for low $q$ and zero
external magnetic field, its dispersion
has been shown to be of the form
$\omega =
\omega_1^+ - \alpha q^2$,\cite{rmk0}
and propagation is possible only in the quantum
hydrodynamic regime (at temperatures for which collisions
are almost absent). This is in contrast to
the Goldstone mode which is purelly quadratic and also
propagates at collision dominated temperatures.\cite{rmk1}
Propagation of the massive mode also requires that the interactions
have some finite amplitude with $p$-wave symmetry.
Lack of observation of such a mode
in neutron scattering experiments
can be attributed to the low spectral weight
the mode carries at small $q$ ($\sim q^2$ when rated
against the Goldstone mode), however indirect signs
of its existence have already been observed.\cite{bla}

We begin our derivation by writing the
equal time propagator for a system
of interacting fermions,
$G_{\sigma_1\sigma_2}({\bf r}_1
{\bf r}_2)\equiv \langle a^\dagger_{\sigma_1}
({\bf r}_1)a_{\sigma_2}({\bf r}_2)\rangle $,
where $a$ is the usual fermion anihilation Heisenberg
operator
and the average is taken with a ferromagnetic ground
state. The equation of motion for $G$ is, in spin space,
$
\partial_tG({\bf r}_1
{\bf r}_2)
+ ({i}/{\hbar })\int d{\bf r'}
[{\cal H}({\bf r}_1{\bf r}')G({\bf r}'{\bf r}_2)
-G({\bf r}_1{\bf r}'){\cal H}({\bf r}'{\bf r}_2)] = g[G],
$
where ${\cal H}$ is the interaction Hamiltonian and the
functional $g[G]$ is the full time-derivative
of $G$.
Following an usual script, one turns to a mixed
${\bf r}$ - ${\bf p}$ representation by defining
${\cal H}_{\bf p}({\bf r}) = \int d{\bf r'}
e^{-i{\bf p}\cdot {\bf r'}/\hbar }
{\cal H}({\bf r}+{\bf r}'/2,{\bf r}-{\bf r'}/2)
$, and likewise for ${\cal N}_{\bf p}({\bf r})$, which
brings the equation of motion to the form
\begin{eqnarray}
\partial_t{\cal N}_{\bf p}({\bf r})
+ \int &&D\tau
\left[{\cal H}_{{\bf p'}+\hbar {\bf k}/2}
({\bf r'}+\hbar {\bf s}/2){\cal N}_{\bf p'}({\bf r'})\right.
\nonumber \\
-&&\left.{\cal N}_{\bf p'}({\bf r'})
{\cal H}_{{\bf p'}-\hbar {\bf k}/2}
({\bf r'}-\hbar {\bf s}/2)\right]
 = J[{\cal N}_{\bf p}],
\label{wig2}
\end{eqnarray}
where $D\tau \equiv id{\bf r'}
d{\bf s}d{\bf p'}d{\bf k}
e^{i[({\bf p'}-{\bf p})\cdot {\bf s} + ({\bf r}-{\bf r'})
\cdot {\bf k}]}/(2\pi )^6\hbar $. Expanding Eq.(\ref{wig2})
in ${\bf s}$
and ${\bf k}$ yields a formal series in $\hbar $,
\begin{eqnarray}
\partial_t{\cal N}_{\bf p}({\bf r})-J&&[{\cal N}_{\bf p}]
=\sum_{n,m=0}^{\infty} \frac{(-1)^n}{2}\left(\frac{i\hbar}{2}\right)^{n+m-1}
\nonumber \\
&&\times \prod_{j=1}^{n}\prod_{l=1}^{m}\frac{\partial^2}{\partial r_{\alpha_j}
\partial p_{\beta_l}}
\left[\frac{\partial^2 
{\cal H}_{\bf p}({\bf r})}{\partial r_{\beta_l}\partial p_{\alpha_j}},
{\cal N}_{\bf p}({\bf r})\right]_{\pm},
\label{wig3}
\end{eqnarray}
where we have (anti) commutation if $m+n$ is
(odd) even and sumation over repeated Greek indices.

It is clear from this expression that the
repeated action of the derivatives on the fluctuating
fields generate coefficients
that are higher order in $q$ for higher powers of $\hbar $.
If we keep terms up to first order in $q$, 
\begin{eqnarray}
\partial_t{\cal N}_{\bf p}({\bf r})+\frac{i}{\hbar }\left[
{\cal H}_{\bf p}({\bf r})\right.,
{\cal N}_{\bf p}&&\left.({\bf r})\right]
-\frac{1}{2}\left\{
{\cal H}_{\bf p}({\bf r}),{\cal N}_{\bf p}({\bf r})\right\}
\nonumber \\
+\frac{1}{2}&&\left\{
{\cal N}_{\bf p}({\bf r}),{\cal H}_{\bf p}({\bf r})\right\}
=J[{\cal N}_{\bf p}],
\label{wig4}
\end{eqnarray}
where square braces indicate the usual commutator
while curly braces are Poisson brackets.
Therefore,
keeping first order in $q$ leads to
the same
result obtained by dropping terms in the series
that are explicitly ${\cal O}(\hbar^2)$. However,
${\cal H}_{\bf p}({\bf r})$
is itself a function of $\hbar $, which disables,
at this point, comparison with a semi-classical
approach.

For the ferromagnetic Fermi liquid there are
two distinct underlying scenarios that we call
(following Ref.\cite{bla}) classical
and quantum spin hydrodynamics.
We want here to discuss in a general
way the origin
of these two regimes. For this purpose,
it should be
recalled that the equilibrium density matrix
for a ferromagnetic state is written in
the quite general form,\cite{abr}
\begin{equation}
{\cal N}^0_{{\bf p}\sigma \sigma'}=n^0_{\bf p}
\delta_{\sigma \sigma'} +
\eta ({\bf p}){\rm \bf m}_0\cdot {\bm \tau}_{\sigma \sigma'},
\label{equilibr}
\end{equation}
where ${\bm \tau}$ are Pauli matrices
and ${\rm \bf m}_0$ is the equilibrium magnetization
density,
which is a conserved quantity of the ferromagnetic
Hamiltonian. From this it is clear that the commutator
in (\ref{wig4})
is zero unless there are fluctuations about equilibrium.

At high temperatures, thermal fluctuations dominate
and this commutator remains irrelevant. 
The remaining ``classical'' terms
lead to the known Bloch-like hydrodynamics
for the Goldstone mode. This regime is refered to
as ``classical
spin hydrodynamics.''

At low temperatures quantum
fluctuations dominate, and the commutator in (\ref{wig4})
becomes important. It remains
finite as long as the fluctuations contain, besides the
usual Goldstone-mode, contributions from
non conserved quantities as it is
the case of spin-current, whose oscillations give rise to
the gapped mode studied in Ref.\cite{bla}, where this regime
has been called ``quantum
spin hydrodynamics.''

The crossover temperature
for these two regimes has been found to scale with $(\omega_1^+)^{1/2}$,
where $\omega_1^+ \equiv 2|{\rm \bf m}_0|(F_1^a/3-F_0^a)/\hbar N(0)$ is the
gap, $F_{\ell}^{(a)s}$ are the usual dimensionless spin (anti) symmetric interaction
parameters, and $N(0)$ is the density of states at the Fermi surface.

The preceding discussion is general and establishes
the origin of the massive mode.
We now turn to the
small moments limit, where kinetic equations
can be amenably derived.
In this limit, we write the effective
quasiparticle Hamiltonian
and density matrix respectively as
$
{\cal H}_{{\bf p}\sigma \sigma'}({\bf r})=
\epsilon_{\bf p}({\bf r})
\delta_{\sigma \sigma'}
+{\bf h}_{\bf p}({\bf r})\cdot
{\bm \tau}_{\sigma \sigma'},
$
and
$
{\cal N}_{{\bf p}\sigma \sigma'}({\bf r})=n_{\bf p}({\bf r})
\delta_{\sigma \sigma'}
+{\bm \sigma}_{\bf p}({\bf r})\cdot
{\bm \tau}_{\sigma \sigma'}
$.
These forms for the Hamiltonian and density immediately
give
\begin{equation}
\frac{i}{\hbar }\left[
{\cal H}_{\bf p}({\bf r}),
{\cal N}_{\bf p}({\bf r})\right] =
\frac{2}{\hbar }\left(
{\bm \sigma}_{\bf p}({\bf r})\times
{\bf h}_{\bf p}({\bf r})
\right)\cdot {\bm \tau}.
\label{commut}
\end{equation}
Here the internal field is given by
\begin{equation}
{\bf h}_{\bf p}({\bf r})=
-\frac {\gamma \hbar }{2}{\bf H}+\frac{2\pi^2 \hbar^2}{m^*k_F}
\sum_{{\bf p'}\ell}F_\ell^a P_\ell({\hat{\bf p}}\cdot{\hat{\bf p}'})
{\bm \sigma}_{\bf p'}({\bf r}),
\label{hp}
\end{equation}
where
$P_\ell $ are Legendre polynomials, ${\bf H}$ is an external
magnetic field,
and all coupling constants are $\hbar $ independent.
We see then from Eqs.(\ref{commut}) and (\ref{hp})
that the commutator of Eq.(\ref{wig4}) gives two
contributions: a zero-th order term in $\hbar $ which
is just the (Larmor) precession of the internal magnetization
and a first order
term in $\hbar $ that is the precession about the
field generated internally. Carefull examination of
Eq.(\ref{wig3})
shows that this term is the $only$ first order
term in $\hbar $ of the series.
That is, provided that the effective
Hamiltonian is of the Fermi liquid type, the long
wavelength limit is equivalent to a
Wentzel-Kramers-Brillouin or (eikonal) semi-classical
approximation.

We want now to have the Lorentz-force
operator explicitly factorized. This is achieved by rewriting
Eq.(\ref{wig4}) in a gauge transformed ``frame''
for which ${\bf A}=0$. Also,
to leading order in the fluctuations, the non-commuting
parts of the Poisson brackets can be dropped, giving
\begin{eqnarray}
\partial_t{\cal N}_{\bf p}({\bf r})+\frac{i}{\hbar }&&\left[
{\cal H}_{\bf p}({\bf r}),
{\cal N}_{\bf p}({\bf r})\right]
+\left\{
{\cal N}_{\bf p}({\bf r}),{\cal H}_{\bf p}({\bf r})\right\}
\nonumber \\
+&&e\left(
\frac{\partial {\cal N}_{\bf p}({\bf r})}
{\partial {\bf p}}\times
\frac{\partial {\cal H}_{\bf p}({\bf r})}{\partial {\bf p}}
\right)\cdot {\bf H}
=J[{\cal N}_{\bf p}].
\label{lin1}
\end{eqnarray}
We see that the gauge boost factorizes a
non-chiral term
whose
amplitude is proportional to $e|{\bf H}|$.
This is the contribution from the orbital
motion of charged quasiparticles, a
result similar to the one found in
the theory of normal metals,\cite{pla}
where it is known simply to shift the
paramagnetic resonances.
In the isotropic {\it ferromagnetic} case we are studying
here this term plays a more essential role. Equation
(\ref{lin1}) is formally identical to the one
obtained in a normal metal. The difference rests
on the broken symmetry ground state, whose density
is given by Eq.(\ref{equilibr}).

Let ${\cal R}^{+(-)}$ be a simultaneous proper
(improper)
rotation of
real and momentum axes and likewise ${\cal SU}^{+(-)}$
for the spin axes. Examination of Eq.(\ref{lin1})
reveals that in the absence of a magnetic field,
${\cal R}^{\pm}$
are symmetry operations both in the paramagnetic and
ferromagnetic cases, ${\cal SU}^{-}$ is not a symmetry
in either case, and ${\cal SU}^{+}$ distinguishes
the two cases, being a broken symmetry in the
ferromagnetic case.
The presence
of ${\bf H}$ adds two terms to the equation, one that couples
to spin through ${\bf h}_{\bf p}$
and the orbital
one discussed above which couples to charge.
In a charged paramagnet, the former will break
${\cal SU}^{+}$ symmetry
yielding propagating paramagnons
as a consequence, while the latter will break
${\cal R}^{-}$ (chiral) symmetry. However,
since the spin modes are {\it sustained} by
a finite ${\bf H}$,\cite{sil}
the degeneracies associated
with chiral symmetry are lifted
with ${\bf H}$, {\it before} spin waves can propagate.

\begin{figure}[h]
\epsfxsize=3.0in
\centerline{\hspace{0.3cm}\epsfbox{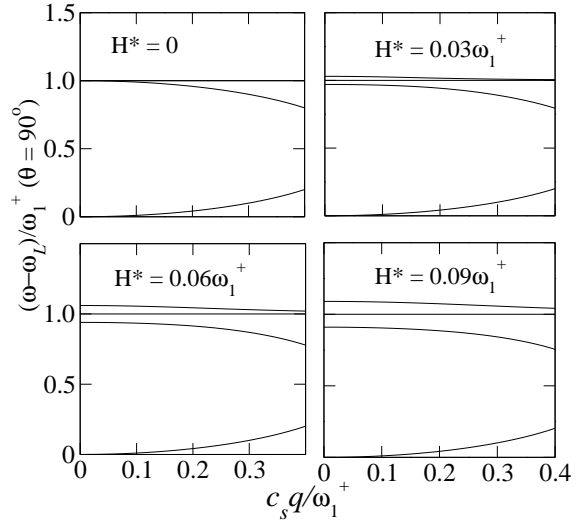}}
\vspace{-3.4truecm}
\renewcommand{\baselinestretch}{1.1}
\caption{Dispersion relations at vanishing temperatures.
${\rm H}^* \equiv
e|{\bf H}_0|(1+F_0^a/3)/m^*$.
\label{fig1}}
\end{figure}

In the isotropic ferromagnet, spin waves can propagate
in the absence of ${\bf H}$, for the ${\cal SU}^{+}$
is a spontaneously broken symmetry and the internal
``sustaining'' field is provided by ${\bf m}_0$.
Hence propagating modes exist before
the formation of orbits with the breaking of
${\cal R}^{-}$
by a finite ${\bf H}$, and some of these modes may
be degenerate.
Breaking chiral symmetry
should then lift such degeneracies.
Note that the Goldstone mode should {\it not}
be affected by ${\bf H}$ (apart, of course, from shifting)
for the term breaking chiral symmetry only couples to
charge. However, the gapped spin-wave mode splits
as it is seen in the dispersion relations
of Fig.(\ref{fig1}).

In order to see how this happens,
we
finish linearizing Eq.(\ref{lin1}), and then
solve it in the hydrodynamic limit.\cite{rmk0}
In particular, since
we are seeking equations on the total
magnetization density
${\bm m}\equiv {\rm tr}[{\bm \tau}
\sum_{\bf p}{\cal N}_{\bf p}]$, and spin
current tensor ${{\bm j}_{\sigma }}_i \equiv {\rm tr}[{\bm \tau}
\sum_{\bf p}(\partial {\cal H}_{\bf p}/\partial p_i){\cal N}_{\bf p}]$,
we trace the product of ${\bm \tau}$ with Eq.(\ref{lin1})
and keep only terms that are linear
in the fluctuations. The result is a set of coupled equations,
\begin{equation}
\partial_t \delta {\bm m}+\nabla_\alpha{{\bm j}_{\sigma }}_\alpha =
-\gamma {\bf H}_0\times \delta {\bm m} - \gamma \delta {\bf H}
\times {\bf m}_0,
\label{mag}
\end{equation}
which is just the continuity equation, and
\begin{eqnarray}
\partial_t {{\bm j}_{\sigma }}_\alpha - c_s^2\nabla_\alpha
&&(\delta {\bm m}+\chi_0\delta {\bf H}) + {\rm H}^*\epsilon_{\alpha z \beta}
{{\bm j}_{\sigma }}_\beta
\nonumber \\
&&= -\gamma ({\bf H}_0
-\omega_1^+ \hat{\bf m}_0)\times {{\bm j}_{\sigma }}_\alpha
-\omega_D^*{{\bm j}_{\sigma }}_\alpha .
\label{cur}
\end{eqnarray}
Here $\delta {\bm m}\equiv {\bm m}-{\bf m}_0$,
$\delta {\bf H}\equiv {\bf H}-{\bf H}_0$ where
${\bf H}_0$ is a uniform magnetic field applied
parallel to the $z$ axis defined by
the equilibrium magnetization ${\bf m}_0$,
$\chi_0\equiv \gamma \hbar N(0)/2|1+F_0^a|$, ${\rm H}^*
=e|{\bf H}_0|(1+F_1^a/3)/m^*$ is the orbital amplitude,
$c_s^2 = (v_F^2/3)|1+F_0^a|(1+F_1^a/3)$ is the squared spin wave
velocity, and $\omega_D^*=(1+F_1^a/3)/\tau_D$
is the $p$-wave amplitude of the scattering integral for a
spin-diffusion relaxation time $\tau_D$.

\begin{figure}[h]
\epsfxsize=3.0in
\centerline{\hspace{0.3cm}\epsfbox{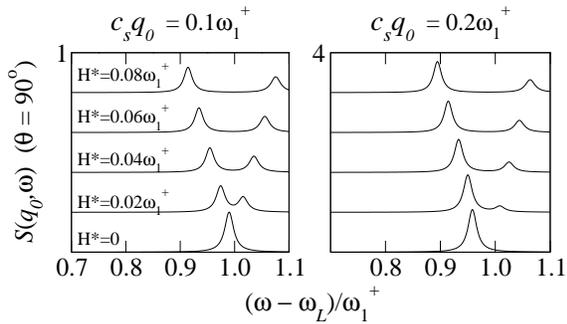}}
\vspace{-6.3truecm}
\renewcommand{\baselinestretch}{1.1}
\caption{Spectral weight in arbitrary units.
The splitting is clearly seen as the magnetic
field increases.
Numbers on the upper left corner give the
amplitude of one plot relative to the next.
These amplitudes
are of order of $10^{-2}$ the (not shown)
Goldstone mode amplitudes which {\it do
not} split under the external field.
The curves are translated vertically to
facilitate analysis.
The values of ${\bf H}^*$ correspond
to hundreds of Gauss in, e.g., MnSi.
\label{fig2}}
\end{figure}

We consider the response to a
driving field $\delta {\bf H}$ which is
transverse to $\hat{z}$ and oscillates
with frequency $\omega $. It is then easy to see that
the three longitudinal components $\hat{z}\cdot {{\bm j}_\sigma }_\alpha $
vanish together with $\delta m_z$. This means that there are no
magnetization gradients in the longitudinal direction and
there is no transport of longitudinal magnetization in any
direction.
The remaining 6 components
can be conveniently ``folded'' onto
the complex ``vector'' ${\bm j}^+_\sigma \equiv \hat{x}\cdot
{{\bm j}_\sigma }_\alpha \hat{x}_\alpha + i\hat{y}\cdot
{{\bm j}_\sigma }_\alpha \hat{x}_\alpha $, which measures
the transport of transverse magnetization $\delta m^+ \equiv
\delta m_x + i\delta m_y$ under the driving field $\delta H^+ \equiv
\delta H_x + i\delta H_y$. Solving Eqs.(\ref{mag}) and (\ref{cur})
for the Fourier components yields\cite{full}
\begin{eqnarray}
{\bm j}^+_\sigma 
= -\frac{\delta H^+c_s^2}{P_4(\Omega)}
[\gamma &&|{\bf m}_0|-\chi_0 (\omega
-\omega_L)]
\nonumber \\
\times &&[\Omega^2{\bf q}-i\Omega {\bf H}^*\times {\bf q}
-({\bf H}^*\cdot {\bf q}) {\bf H}^*],
\label{jplus}
\end{eqnarray}
and
$
{\delta m}^+ 
= -{\delta H}^+{\gamma |{\bf m}_0| P_3(\Omega )}/{P_4(\Omega )},
$
where $\Omega \equiv \omega - \omega_L - z_0$,
$\omega_L$ is the Larmor frequency,
$z_0\equiv \omega_1^+ - i\omega_D^*$,
$
P_3(\Omega )=\Omega^3+(c_s^2 q^2{\chi_0}/{\gamma |{\bf m}_0|})
\Omega^2 - {{\rm H}^*}^2\Omega - 
(c_s^2{\chi_0}/{\gamma |{\bf m}_0|})
({\bf H}^*\cdot {\bf q})^2,
$
$
P_4(\Omega )=\Omega^4+z_0\Omega^3 +
(c_s^2q^2-{{\rm H}^*}^2)\Omega^2 -z_0
{{\rm H}^*}^2\Omega -c_s^2({\bf H}^*\cdot {\bf q})^2,
$
and
${\bf H}^*\equiv {\rm H}^*\hat{z}$.
For simplicity
we specialize to the case ${\bf H}_0 \perp
{\bf q}$ (indicated by $\theta =90^o$
in the figures). This condition is obtained in practice
by sheding neutrons or photons adequately
on a finite sample's surface, and is the condition
that maximizes the non-chiral orbital contribution
in Eq.(\ref{jplus}); in the infinite
system, we can think of transverse magnons
if we wish to quantize the hydrodynamics.
Longitudinal magnons will be treated elsewhere.\cite{full}
In Fig.(\ref{fig1}) we see the
dispersion relations for different values of the external
field. These are obtained by looking into the free modes
($\delta {\bf H}=0$). 
It is clear that the degeneracy is only exact
at $q=0$.
We see also that it constitutes a three-fold
degeneracy, however, the two constant
branches shown ($\omega - \omega_L \equiv
\Delta \omega = \omega_1^+$) are spurious
when ${\bf H}_0=0$ in the
sense that there is no spectral weight associated with them,
as seen in Fig.(\ref{fig2}) where $S(q,\omega)\propto
{\rm Im}[\delta m^+/\delta H^+]$ (which is the important quantity
in experiments)
is plotted as a function of $\omega $ and ${\rm H}^*$.
Under a finite magnetic field only one
of the spurious branches develops weight while the
degeneracy is lifted.

\begin{figure}[h]
\epsfxsize=3.0in
\centerline{\hspace{0.3cm}\epsfbox{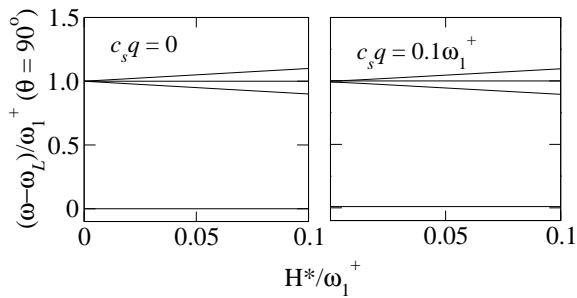}}
\vspace{-6.4truecm}
\renewcommand{\baselinestretch}{1.1}
\caption{Linear behavior of the dispersions with ${\rm H}^*$.
The effect of ${\rm H}^*$ on the
Goldstone mode ($\omega - \omega_L\sim
0$)
is seen to merely shift its
frequency.
\label{fig3}}
\end{figure}

The important point to
note is that the frequencies of the
Goldstone mode (the traditional
spin-waves) only shift their loci as the external
magnetic field
increases whereas the gapped mode splits. This can be seen in
Fig.(\ref{fig3}) where the behavior of these
dispersions with the external field is shown
(including the Goldstone mode).
It is then clear that the splitting of the
gapped mode
should be considered as a distinctive
feature in experiments searching for a direct
observation of it.
It should be helpful to make a statement
about the quantities involved for some traditional
materials: From
data found in the literature
we estimate
the fields shown
to be in the range of hundreds of Gauss for typical weak
ferromagnets like MnSi, ZrZn$_2$, and Ni$_3$Al. This corresponds to a
gap of tenths of meV in these materials.\cite{ishi}

For small values of ${\rm H}^*$ and $q$ (up to $0.1\omega_1^+$),
the dispersions may be put in the simple form
$\Delta \omega = c_s^2q^2/\omega_1^+$ for the
Goldstone mode and $\Delta \omega = \pm {\rm H}^* + \omega_1^+
- \alpha q^2$ for the gapped mode.\cite{rmk7}
Quantization of the theory in this
limit will yield massive magnons with an ``up-down'' degeneracy
which is lifted by the magnetic field, in much the same way
as in the Zeeman effect of a single spin.
The possible existence of (low temperature) magnons
with mass and a Zeeman-like degeneracy makes room for new
questions realated, e.g., to the collective behavior
of a handful of these excitations. Before these questions
are put forward it is, however, prudent to focus on the
issue of whether these results can be observed by traditional
experiments. The spectral
weight of this mode, of order of $10^{-2}$ the usual spin waves
(see Fig.(\ref{fig2})) stands
as an obstacle. Our main emphasis here is whence on the additional
criterium such a degeneracy provides to track down the massive mode.
In a companion longer article we also discuss how
these results are expected to show in a conduction ressonance experiment.
\cite{full}

The authors aknowledge fruitful discussions with K. Blagoev.
Financial support for this work
has been provided by DOE Grant DEFG0297ER45636, and
FAPESP Grants 01/01713-3, 00/07660-6, and 00/10805-6.


\begin{references}
\bibitem{rmk0} In this article
the so called hydrodynamic approximation is used which
consists of keeping only the two first spherical harmonics
in the expansions of the responses about the Fermi surface.
To properly determine $\alpha $ one has to go beyond
such an approximation , as in Ref.\cite{bla}, where it has
been found (in the absence of an external field)
that $\alpha = c_s^2/\omega_1^+ - 6n/5F_1^am^*m^2$
($n$ is the density, $m$ and $m^*$ the bare and effective masses and
$F_1^a$ the second Landau antisymmetric parameter).
\bibitem{rmk1} The fact that the gapped mode does not propagate in the
presence of collisions is due to its origin in the
oscillations of spin-current, which is not a conserved
quantity.\cite{bla}
\bibitem{rmk7} For the strict hydrodynamic approximations used here, $\alpha
= c_s^2/\omega_1^+$. A more rigorous expression for
the relation between $\alpha $ and the Landau parameters in a
finite magnetic field will be presented in the future;
we do not expect $\alpha $ to change, to leading order in the field,
from the expression derived in Ref.\cite{bla}.\cite{rmk0}
\end{references}
\end{document}